\documentclass[twocolumn]{aastex61}
\usepackage{amsmath}
\received{\today}
\revised{\today}
\accepted{\today}

\submitjournal{ApJL}

\shorttitle{Beyond UVJ}
\shortauthors{Leja, Tacchella, \& Conroy}

\begin{document}

\title{Beyond UVJ: \\ More Efficient Selection of Quiescent Galaxies With UV / Mid-IR Fluxes} 

\email{joel.leja@cfa.harvard.edu, sandro.tacchella@cfa.harvard.edu}

\author[0000-0001-6755-1315]{Joel Leja}
\affil{Harvard-Smithsonian Center for Astrophysics, 60 Garden St. Cambridge, MA 02138, USA}

\author[0000-0002-8224-4505]{Sandro Tacchella}
\affil{Harvard-Smithsonian Center for Astrophysics, 60 Garden St. Cambridge, MA 02138, USA}

\author[0000-0002-1590-8551]{Charlie Conroy}
\affil{Harvard-Smithsonian Center for Astrophysics, 60 Garden St. Cambridge, MA 02138, USA}

\begin{abstract}
    The UVJ color-color diagram is a popular and efficient method to distinguish between quiescent and star-forming galaxies through their rest-frame $U-V$ vs. $V-J$ colors. Here we explore the information content of this color-color space using the Bayesian inference machine \texttt{Prospector}. We fit the same physical model to two datasets: (i) UVJ fluxes alone, and (ii) full UV-mid IR (MIR) broadband SEDs from the 3D-HST survey. Notably this model uses both nonparametric SFHs and a flexible dust attenuation curve, both of which have the potential to `break' the typical correlations observed in UVJ color-color space. Instead, these fits confirm observed trends between UVJ colors and observed galaxy properties, including specific star formation rate (sSFR), dust attenuation, stellar age, and stellar metallicity. They also demonstrate that UVJ colors do not, on their own, constrain stellar age or metallicity; the observed trends in the UVJ diagram are instead driven by galaxy scaling relationships and thus will evolve with cosmological time. We also show that UVJ colors ``saturate'' below $\log(\mathrm{sSFR/yr}^{-1})\lesssim -10.5$, i.e. changing sSFR no longer produces substantial changes in UVJ colors. We show that far-UV and/or MIR fluxes continue to correlate with sSFR down to low sSFRs and can be used in color-color diagrams to efficiently target galaxies with much lower levels of ongoing star formation. We provide selection criteria in these new color-color spaces as a function of desired sample sSFR.
\end{abstract}


\section{Introduction}
Quantifying the rate of stellar mass assembly in star-forming and quiescent galaxies over the past $\sim13$ Gyr is necessary to understand how the present-day galaxy population formed. Such an investigation requires large numbers of both star-forming and quiescent galaxies at early epochs. Separating the two populations can be challenging, with the most direct methods requiring expensive spectroscopic measurements such as D$_{\mathrm{n}}$4000 ({Kauffmann} {et~al.} 2003) or H${\alpha}$ equivalent width ({Brinchmann} {et~al.} 2004). 

During the last decade, rest-frame color-color diagrams and in particular the UVJ diagram have been very popular for separating these two categories of galaxies, in part because they can be efficiently applied to large photometric samples (e.g., {Daddi} {et~al.} 2004; {Williams} {et~al.} 2009; {Arnouts} {et~al.} 2013). {Williams} {et~al.} (2009) originally devised the UVJ color-color selection, based on the corresponding color-color diagram introduced by {Wuyts} {et~al.} (2007) (see also BzK selection, {Daddi} {et~al.} 2004). This approach uses near-infrared photometry to solve the long-running problem of distinguishing between galaxies which are optically red due to age and galaxies which are optically red due to dust attenuation (e.g., {Strateva} {et~al.} 2001; {Balogh} {et~al.} 2004; {Baldry} {et~al.} 2004. Since then, UVJ selection has been used to sort galaxy samples at all cosmic epochs with great success (e.g., {Whitaker} {et~al.} 2013; {Barro} {et~al.} 2014; {Straatman} {et~al.} 2014; {Papovich} {et~al.} 2018). The efficacy of this selection has been confirmed with deep MIR imaging revealing low average sSFRs in UVJ-selected quiescent galaxies: sSFR $\sim10^{-11.9} \times (1 + z)^4$ $\mathrm{yr}^{-1}$ ({Fumagalli} {et~al.} 2014). Simulations have even begun assigning UVJ colors to their outputs in order to define quiescence (e.g., {Dav{\'e}}, {Rafieferantsoa}, \&  {Thompson} 2017; {Donnari} {et~al.} 2018). 

However, advances in statistics, modeling, and reams of new data have provided sophisticated tools to evaluate UVJ classification in new detail. Recent studies using spectroscopic information ({Belli} {et~al.} 2017; {Schreiber} {et~al.} 2018), spectral energy distribution (SED) fitting ({D{\'{\i}}az-Garc{\'{\i}}a}  {et~al.} 2017; {Fang} {et~al.} 2018; {Merlin} {et~al.} 2018), and combinations of methods ({Moresco} {et~al.} 2013) find that UVJ-quiescent selection includes $\sim10-30\%$ contamination from star-forming galaxies. Furthermore, there exist correlations in the quiescent part of the UVJ diagram which permit measurements of ages ({Whitaker} {et~al.} 2013; {Belli}, {Newman}, \& {Ellis} 2018), and when UVJ colors are combined with stellar mass and redshift, it has been claimed that one can measure metallicities, extinctions, and sSFRs as well ({D{\'{\i}}az-Garc{\'{\i}}a}  {et~al.} 2017).

Building on these findings, here we use the Bayesian inference machine \texttt{Prospector} (Johnson \& Leja 2017; {Leja} {et~al.} 2017) to examine the ability of straightforward UVJ color-color cuts to diagnose stellar populations properties. Bayesian inference is the natural tool for this task as it is designed to deal with complex correlations such as those that exist between galaxy properties and rest-frame colors.

Throughout the paper we use a {Chabrier} (2003) IMF and a WMAP9 ({Hinshaw} {et~al.} 2013) cosmology. All parameters are reported as the median of their respective probability distribution and all magnitudes are in the AB system.

\section{Data and Models}
We use the \texttt{Prospector} Bayesian inference machine ({Leja} {et~al.} 2017; Johnson \& Leja 2017) to translate galaxy photometry into parameter posteriors. This approach uses the Flexible Stellar Population Synthesis (FSPS) stellar populations ({Conroy}, {Gunn}, \& {White} 2009) code to construct a physical model and the nested sampler \texttt{dynesty} ({Speagle} 2019) to sample the posterior space. 

\begin{figure*}[h!t]
\begin{center}
\includegraphics[width=0.95\linewidth]{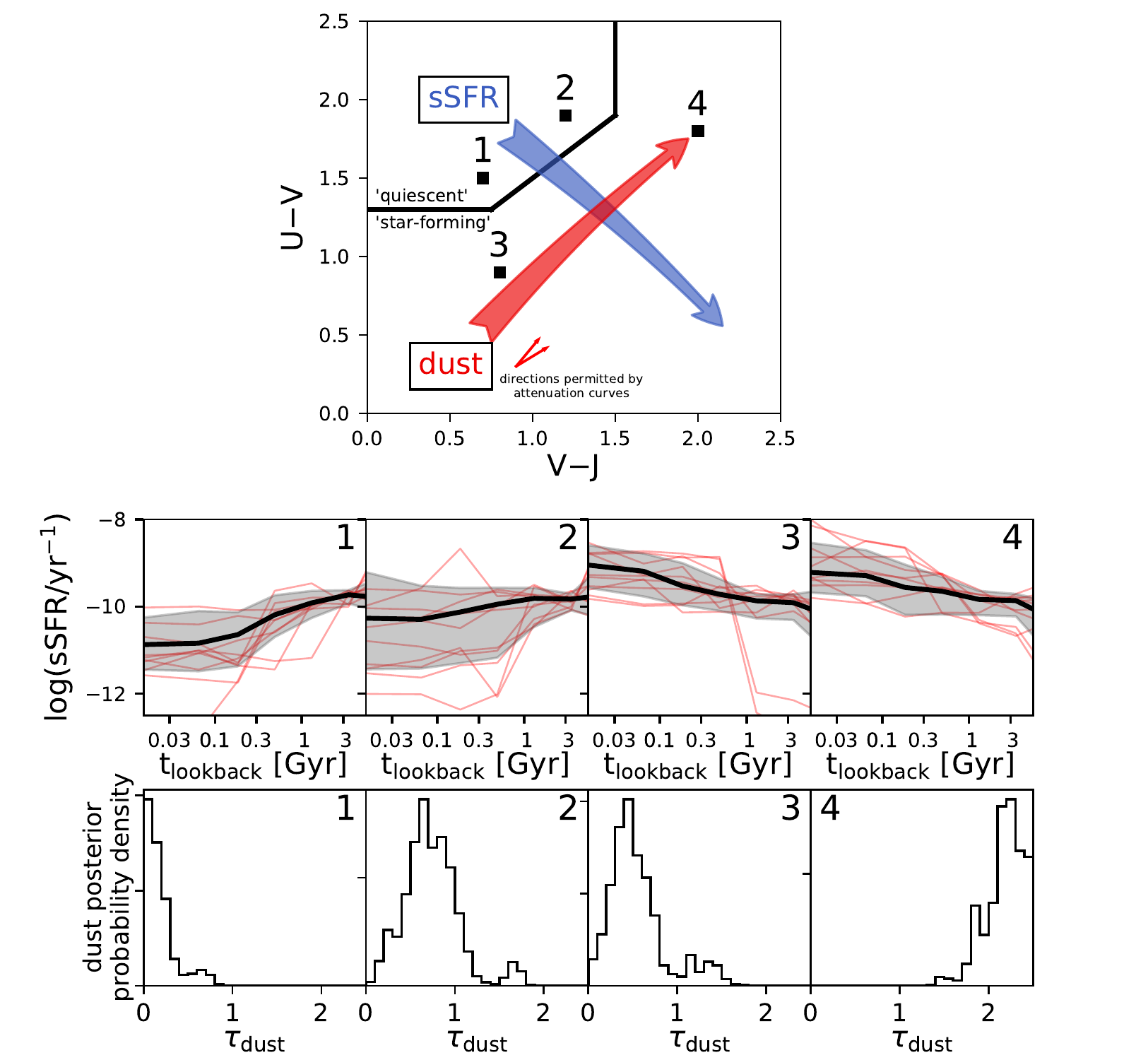}
\caption{The top panel shows the UVJ diagram. The blue and red arrows show the effect of increasing sSFR and dust attenuation, respectively. The black arrows show the range of dust vector angles permitted by changes in the attenuation curve. The lines indicate the UVJ-quiescent selection box. The middle and lower panels show SFH and dust posteriors from fits to synthetic UVJ fluxes corresponding to the numbers in the top panel. For the SFHs, the black line is the posterior median, the grey shaded region is the 1$\sigma$ posterior, and thin red lines are random draws from the posterior. There is a clear mapping from UVJ colors to dust attenuation and the sSFRs of star-forming galaxies, but the quiescent region permits a wide range of recent sSFRs.}
\label{fig:UVJ}
\end{center}
\end{figure*}
Within the \texttt{Prospector} framework we construct two closely related physical models, one optimized to fit observed panchromatic galaxy SEDs and one to fit synthetic UVJ fluxes.
\subsection{Fitting Observed Panchromatic Photometry}
\label{sec:3dhst}
To better understand how the properties of observed galaxies correlate with their rest-frame UVJ colors, we take the physical parameters derived from \texttt{Prospector} fits to the 3D-HST photometric catalogs ({Skelton} {et~al.} 2014) from {Leja} {et~al.} (2019b).

These fits use a modified version of the Prospector$-\alpha$ physical model, described in detail in {Leja} {et~al.} (2019b). In brief, this model includes a 7-bin nonparametric star formation history (SFH) with a prior emphasizing smoothness in SFR(t) ({Leja} {et~al.} 2019a), a two-component dust model with a flexible attenuation curve ({Charlot} \& {Fall} 2000), free stellar metallicity with a mass-metallicity prior, and hot dust emission from an active galactic nucleus ({Leja} {et~al.} 2018). This model also includes dust emission via energy balance and nebular emission self-consistently powered by the stellar fluxes.

The 3D-HST catalogs provide observed-frame 0.3$\mu$m$-$24$\mu$m photometry and redshifts for some 200,000 galaxies. These galaxies are in five well-studied extragalactic fields and are imaged in $19-45$ photometric bands. In this work we use a sub-sample of galaxies with stellar mass M$_* > 10^{10}$ M$_{\odot}$ in the redshift range $0.5 < z < 2.5$, corresponding to 12,235 galaxies.
\begin{figure*}
\begin{center}
\includegraphics[width=0.95\linewidth]{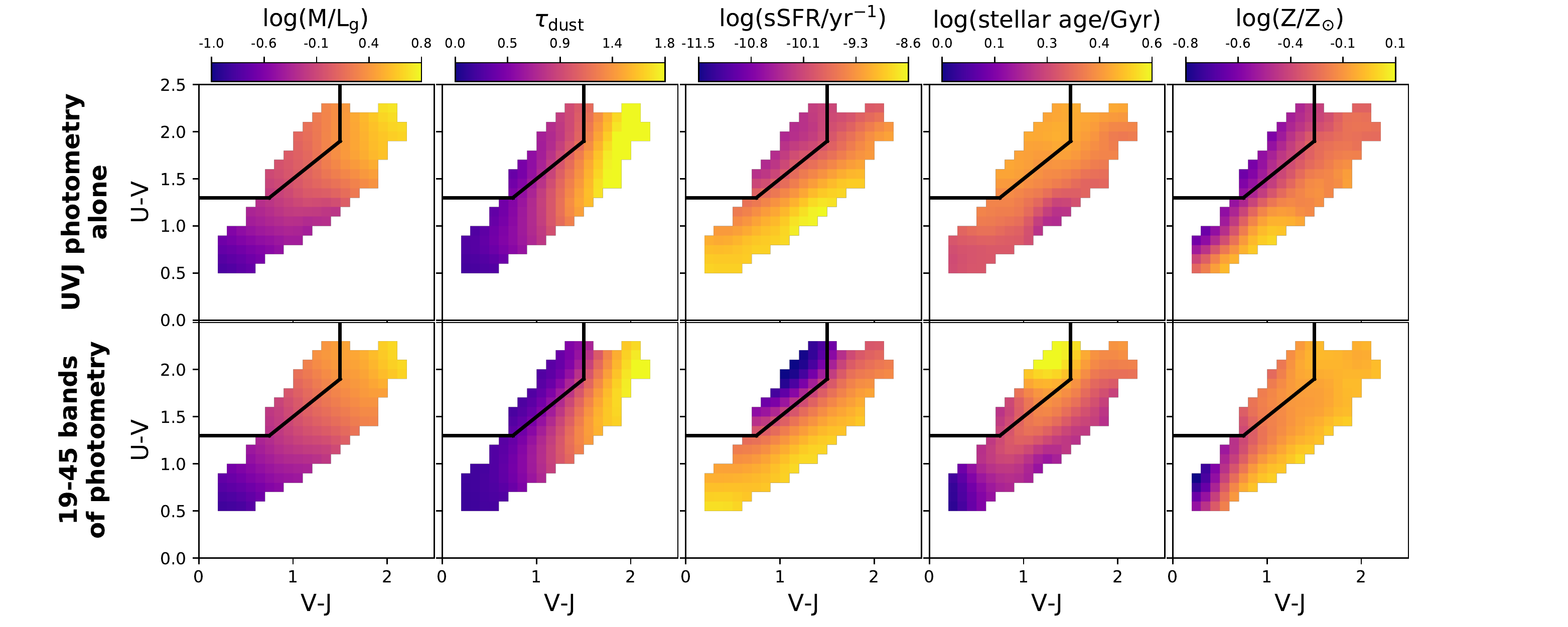}
\caption{Median stellar population properties in the UVJ diagram after fitting synthetic UVJ fluxes (upper panels) and after fitting full SEDs of observed galaxies (lower panels). From right to left: mass-to-light ratio for the SDSS-$g$ band relative to solar, dust optical depth, sSFR averaged over the most recent 100 Myr, average stellar age, and stellar metallicity. Each pixel shows either the median of the posterior (top rows) or the median parameter for galaxies in the UVJ pixel (bottom rows). Constraints from synthetic UVJ fluxes produce strong trends in dust and M/L$_g$ and weak trends in sSFR, metallicity, and age. Comparatively, observed galaxies show stronger trends in sSFR and age and slightly stronger trends in metallicity. This difference implies that galaxies occupy a lower-dimensional parameter space than permitted by their UVJ colors alone.}
\label{fig:parameter_maps}
\end{center}
\end{figure*}

\begin{figure*}
\begin{center}
\includegraphics[width=0.95\linewidth]{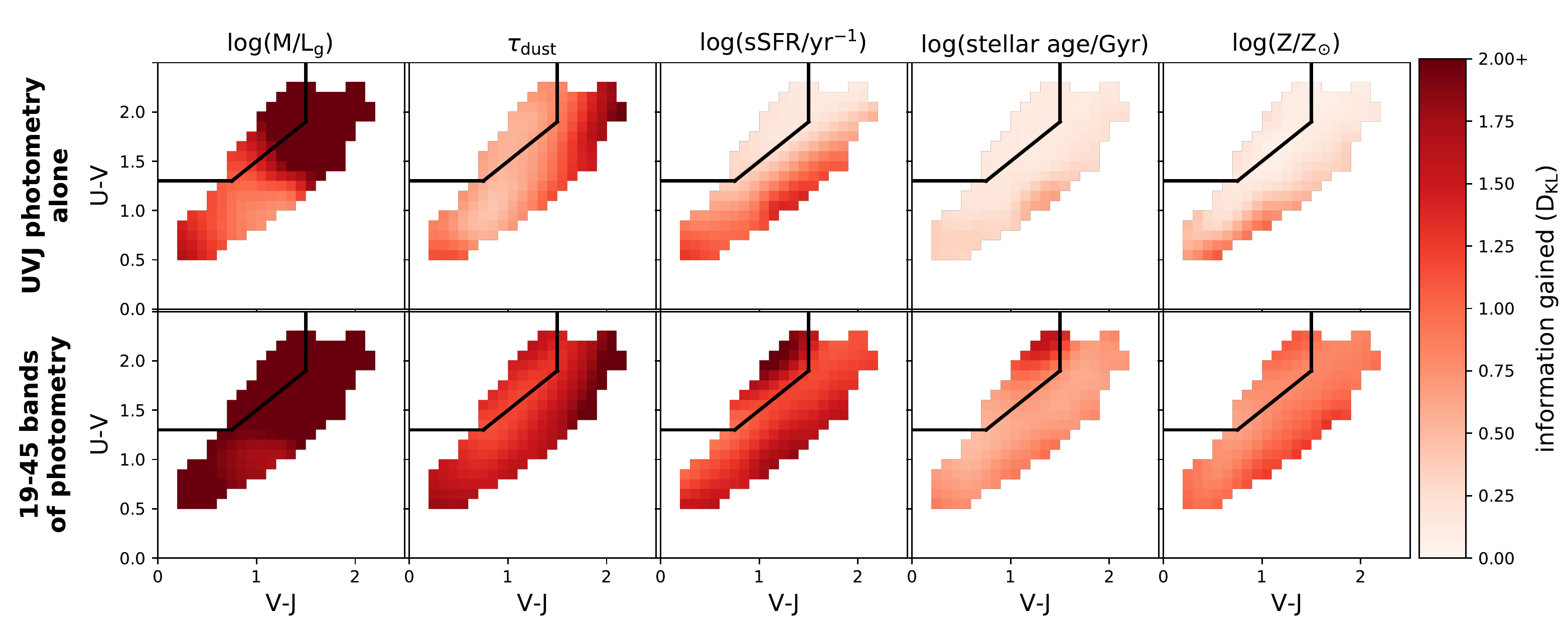}
\caption{Information gained after fitting synthetic UVJ fluxes (upper panels) and full observed SEDs (lower panels). The overall layout of the figure follows Figure~\ref{fig:parameter_maps}. We quantify the information gained from the data by calculating the Kullback-Leibler divergence (D$_{\mathrm{KL}}$) between the prior and the posterior. We find that synthetic UVJ fluxes put tight constraints on M/L$_{\mathrm{g}}$ and dust attenuation, partial constraints on sSFR, and minimal constraints on metallicity and age. The lower panels demonstrate that full SED fits puts more meaningful constraints on these parameters, confirming that the trends in the bottom panels of Figure \ref{fig:parameter_maps} are data-driven rather than a consequence of model assumptions.}
\label{fig:KLD}
\end{center}
\end{figure*}

\subsection{Fitting Synthetic UVJ Fluxes}
\label{sec:uvj_fluxes}
We also fit a grid of rest-frame $U$, $V$, and $J$ fluxes to determine the constraining power of UVJ fluxes alone. These fluxes specify a single UVJ color and are given an arbitrary normalization. We generate 625 sets of UVJ fluxes corresponding to a regular grid in $0 < U-V < 2.5$ and $0 < V-J < 2.5$.

We fit these fluxes with the modified Prospector$-\alpha$ model described above. The mass-metallicity prior is replaced with a flat metallicity prior over $-1.0 < \log(\mathrm{Z/Z}_{\odot}) < 0.2$ to ensure the analysis is independent of stellar mass. The maximum stellar age is set to 6 Gyr, corresponding to the age of the Universe at $z=1$. The fluxes are assigned errors of 2.5\%, though to preserve the UVJ colors the fluxes themselves are not perturbed.

\section{Galaxy Properties in the UVJ Diagram}
{Williams} {et~al.} (2009) show that UVJ selection can separate dusty star-forming galaxies from quiescent galaxies because dusty star-forming galaxies are red in $V-J$ while quiescent galaxies are blue in $V-J$. While largely an empirical finding, this behavior was shown to be consistent with constrained dust models using fixed attenuation curves and parametric SFHs. However, there is a growing body of evidence suggesting that galaxies have a diversity of dust attenuation curves ({Salmon} {et~al.} 2016; {Leja} {et~al.} 2017; {Salim}, {Boquien}, \& {Lee} 2018; {Narayanan} {et~al.} 2018) and a diversity of star formation histories ({Pacifici} {et~al.} 2016; {Iyer} {et~al.} 2019). Here, we use a more complex two-component dust model allowing variation in the shape of the dust attenuation curve and a flexible nonparametric distribution of stellar ages, allowing us to test the robustness of these conclusions to these assumptions.

\begin{figure*}
\begin{center}
\includegraphics[width=0.8\linewidth]{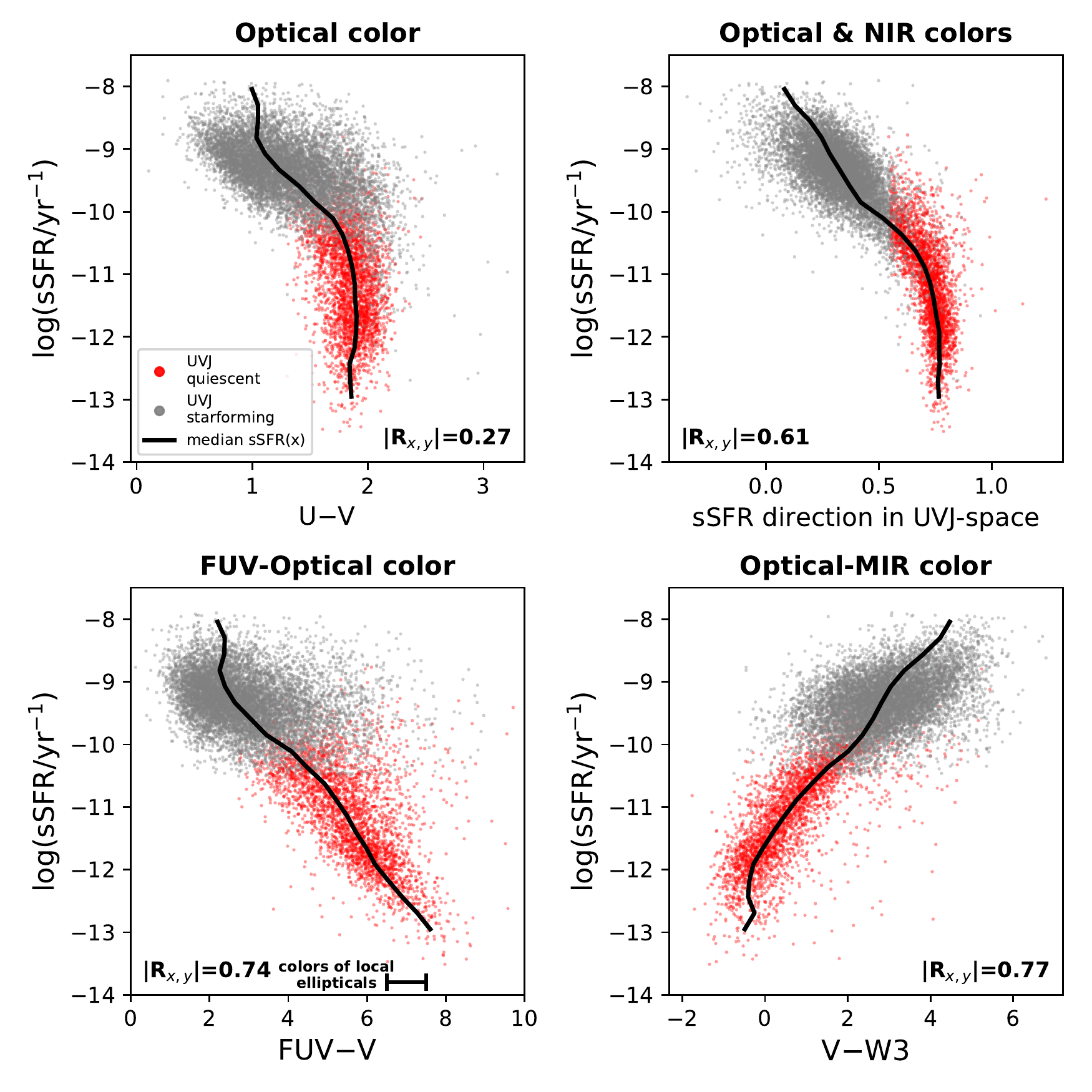}
\caption{The correlation between sSFR and optical color, UVJ colors, UV-optical color and optical-MIR color is shown. The Pearson correlation coefficient between color and sSFR is shown in the corner. Optical and NIR colors begin to saturate at approximately sSFR $\sim 10^{-10.5}$ yr$^{-1}$ whereas FUV-optical or optical-MIR colors continue to correlate with sSFR over a wide range of values. The median sSFR as a function of color is shown as a black line. The approximate range of colors for local elliptical galaxies is shown as a black bar in the $FUV-V$ diagram ({Jeong} {et~al.} 2009), described in the text.}
\label{fig:color_cuts}
\end{center}
\end{figure*}
Figure \ref{fig:UVJ} shows how the SFH and dust posteriors change as a function of rest-frame UVJ colors\footnote{Here we adopt the UVJ quiescent selection criteria from {Whitaker} {et~al.} (2012)}. The posteriors are derived by fitting synthetic UVJ fluxes described in Section \ref{sec:uvj_fluxes}. This illustrates that UVJ colors continue to put robust constraints on both sSFR and dust attenuation even after allowing for the presence of confounding effects like age, stellar metallicity, and flexible dust models.

Figure \ref{fig:parameter_maps} explores further trends between galaxy properties and UVJ colors. Galaxy properties are inferred both from synthetic UVJ fluxes and from fits to the observed UV-MIR SEDs. For the observed galaxies where many objects fall within a single UVJ pixel, the median value is shown. Only pixels containing at least 10 galaxies are shown. The maps are smoothed with a Gaussian with $\sigma$=1 pixel in order to highlight trends.

Some of these parameters, such as M/L$_g$ and dust attenuation, show strong and consistent trends whether fitting simple UVJ fluxes or the full photometric SED. These parameters are well constrained by UVJ colors alone. Other parameters, such as mean stellar age and metallicity, show no structure in UVJ space until constrained by the full photometric SED. These properties are either weakly constrained or not constrained by UVJ colors alone. This suggests that trends in the observed galaxies are induced by galaxy scaling relationships.

sSFR is a special case in this comparison. While the median sSFR of star-forming galaxies is unchanged when constrained with UVJ fluxes or the full SED, the median sSFR of quiescent galaxies becomes much lower. This is because galaxies with moderate sSFRs, e.g. sSFR $\sim10^{-10.5}$ yr$^{-1}$, can also fall into the UVJ-quiescent region. This is a key result of this Letter, discussed further in Section \ref{sec:saturation}.

One important caveat is that it is not clear from the parameter maps alone whether these trends are being driven by the data or by assumptions built into the model. One way to distinguish between the two is to measure the difference between the prior and the posterior distributions. A reliable metric for this is the Kullback-Leibler divergence (hereafter D$_{\mathrm{KL}}$), defined as
\begin{equation}
\mathrm{D}_{\mathrm{KL}} = \int^{\infty}_{-\infty} a(x) \ln\left(\frac{a(x)}{b(x)}\right) dx
\end{equation}
\noindent for two probability distributions $a(x)$ and $b(x)$.

In Bayesian analysis, D$_{\mathrm{KL}}$ calculated from the prior $a(x)$ to the posterior $b(x)$ is interpreted as the information gained by fitting the data. If no information is gained, the prior and the posterior are identical and D$_{\mathrm{KL}}=0$. As D$_{\mathrm{KL}}$ increases, the posterior and the prior become increasingly divergent.

Figure \ref{fig:KLD} is constructed in an analogous fashion to Figure \ref{fig:parameter_maps} and shows D$_{\mathrm{KL}}$ from the prior to the posterior. The median D$_{\mathrm{KL}}$ for the galaxies in the pixel is shown for the full SED fits. The D$_{\mathrm{KL}}$ maps for the synthetic UVJ fluxes confirm our previous conclusions: M/L$_g$, dust attenuation, and the sSFR of star-forming galaxies are fairly well-constrained by UVJ fluxes alone, while ages, metallicities, and the sSFRs of galaxies in the quiescent box are relatively unconstrained. 

The average D$_{\mathrm{KL}}$ increases substantially when fitting the full photometric SED. This is expected, as the full SED provides more information than UVJ fluxes alone. However this provides necessary confirmation that UV-MIR photometry can put meaningful constraints on these parameters and implies that the trends observed in Figure \ref{fig:parameter_maps} are reliable -- though fitting spectroscopic data has the potential to provide much more precise measurements (e.g. {Belli} {et~al.} 2018).

These results taken together imply that the age and metallicity trends in the UVJ diagram are not specified by UVJ colors alone. Instead, galaxies must exist on some constrained plane in a high-dimensional parameter space (i.e. are subject to galaxy scaling relationships), with the shape of this plane then inducing correlations with UVJ colors. This means that these relationships can evolve with cosmological time. Indeed this evolution can be seen directly in the data: for example, the age-color trend in the quiescent box in Figure \ref{fig:parameter_maps} is a combination of a mild age gradient at fixed redshift and the net evolution of the UVJ colors of the galaxy population across redshifts. Also, galaxies with sub-solar metallicity take about 3 Gyr to age into the UVJ-quiescent region ({Tacchella} {et~al.} 2018), implying that UVJ selection will fail to identify low-metallicity quiescent galaxies at $z\ga3$.

We note that D$_{\mathrm{KL}}$ is not invariant to the chosen model. For example, the UVJ-quiescent region has a low D$_{\mathrm{KL}}$ for sSFR when constrained only by UVJ colors. This is partly because UVJ colors are not correlated with sSFR for sSFR $< 10^{-10.5}$ yr$^{-1}$, but also partly because the model is able to produce star-forming galaxies (sSFR$>10^{-9}$ yr$^{-1}$) with UVJ-quiescent colors by combining significant dust attenuation with steep, SMC-like attenuation curves. Such galaxies likely do not exist in the real Universe due to the physical correlation between increasingly flat attenuation curves and increasing dust attenuation (e.g. {Chevallard} {et~al.} 2013): however, they are not ruled out {\it a priori} by the model, resulting in a lower-than-expected D$_{\mathrm{KL}}$. The full SED fits are robust to this effect as the full SED can reliably rule out the combination of high dust attenuation and steep attenuation curves.

\begin{figure*}
\begin{center}
\includegraphics[width=0.95\linewidth]{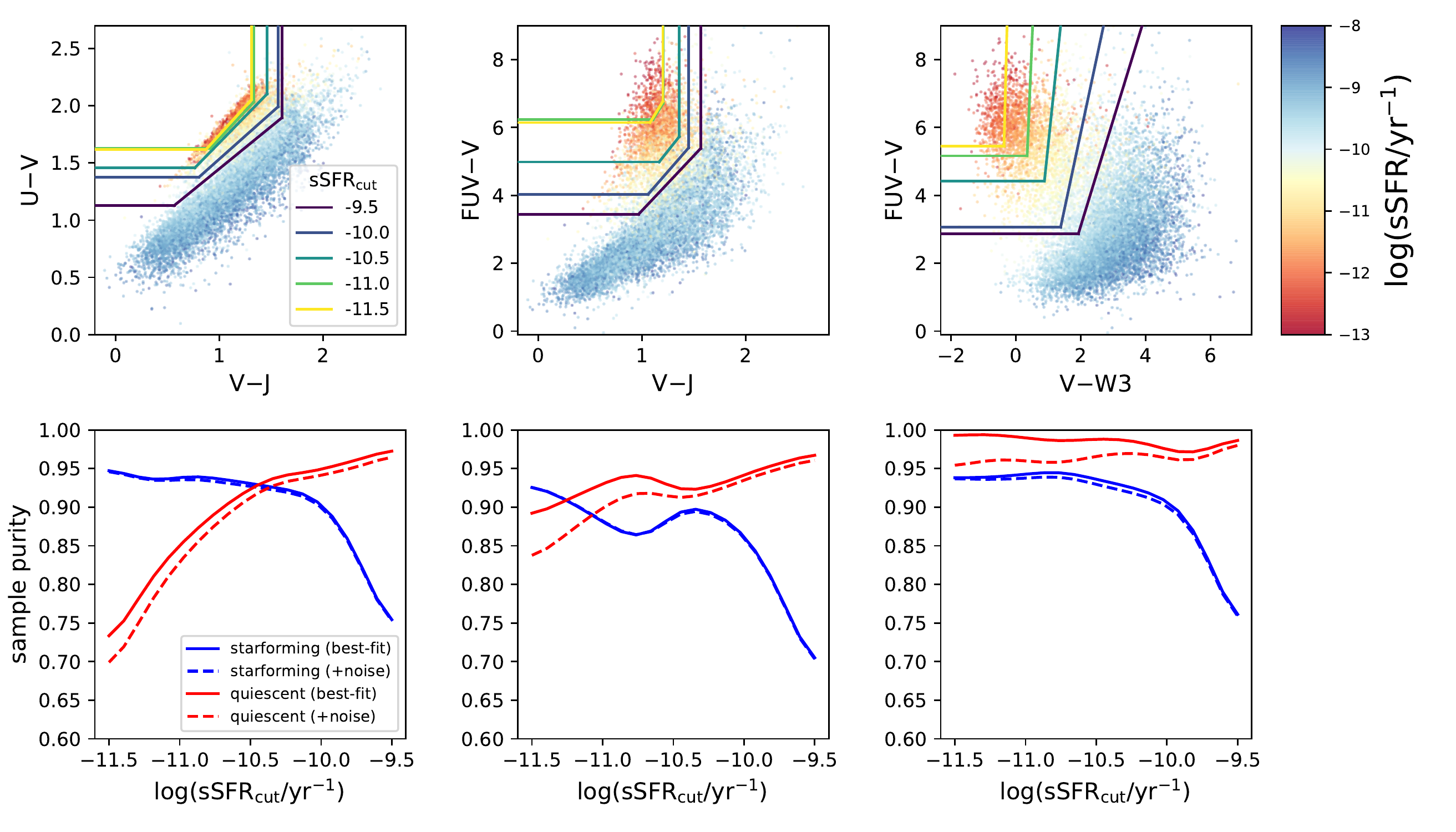}
\caption{Comparing the efficiency of sample selection by galaxy sSFR in different color-color spaces. In the upper row, from left to right, the panels show the canonical UVJ diagram, the FUV-V-J diagram, and the FUV-V-W3 diagram. In the lower row, the sample purity as a function of target sSFR is shown, both for the best-fitting colors (solid line) and modeling the effect of photometric uncertainty (dashed lines). All of the color-color diagrams perform well at sSFR$_{\mathrm{cut}} \sim 10^{-10}$ yr$^{-1}$, but for lower sSFRs the UVJ diagram becomes increasingly inefficient while the FUV/MIR color-color diagrams remain near 100\% purity.}
\label{fig:alternate_UVJ}
\end{center}
\end{figure*}

\section{Beyond UVJ}
\label{sec:saturation}

The fact that quiescent UVJ colors from the quiescent region do not appear to specify low sSFRs merits further investigation. While the UVJ diagram was designed to distinguish between star-forming and quiescent galaxies, Figures \ref{fig:parameter_maps} and \ref{fig:KLD} together suggest that the UVJ quiescent selection cannot distinguish between moderate and low sSFRs. This is consistent with observational findings that $\sim10-30$\% of UVJ-quiescent galaxies host significant ongoing star formation ({D{\'{\i}}az-Garc{\'{\i}}a}  {et~al.} 2017; {Belli} {et~al.} 2017; {Schreiber} {et~al.} 2018).

Figure \ref{fig:color_cuts} examines the correlation between rest-frame colors and quiescence directly by plotting the relationship between several rest-frame colors and $\log(\mathrm{sSFR/yr}^{-1})$ inferred from the \texttt{Prospector} fits to the 3D-HST photometry. The sSFR direction in UVJ-space is defined as the perpendicular direction to the quiescent selection line ({Fang} {et~al.} 2018), indicated by the blue arrow in Figure \ref{fig:UVJ}.

This figure demonstrates the well-known fact that $U-V$ colors alone cannot distinguish between star-forming and quiescent galaxies (e.g., {Eales} {et~al.} 2017). It further shows that while UVJ colors partially break this degeneracy, the correlation between color and sSFR begins to saturate at sSFR $< 10^{-10.5}$ yr$^{-1}$ and is fully saturated by sSFR $= 10^{-11}$ yr$^{-1}$.

The lower panels show how this relationship can be restored by instead using colors calculated with far-UV (FUV) or mid-infrared (MIR) fluxes, i.e. {\it GALEX} FUV ($\lambda_{\mathrm{rest}} \sim 1500$ $\mathrm{\AA}$) and $WISE$ W3 ($\lambda_{\mathrm{rest}} \sim 12\mu$m). Colors constructed with these fluxes correlate with sSFR down to low sSFRs. The correlation between rest-frame color and sSFR is calculated using the Pearson correlation coefficient, shown in the corner of each panel. The increasing coefficient suggests that quiescent galaxies can be more cleanly identified with FUV or MIR fluxes. We note that the outliers in the V-W3 -- sSFR relationship are almost entirely galaxies with significant mid-IR AGN emission: removing such galaxies using AGN indicators such as X-ray luminosities will further increase the efficacy of this selection.

One concern is that hot evolved stars can produce very blue colors which masquerade as star formation in the FUV (e.g., {Han}, {Podsiadlowski}, \& {Lynas-Gray} 2007). We include observed $FUV-V$ colors from local quiescent galaxies as a rough upper bound for the size of this effect ({Jeong} {et~al.} 2009). Notably, the abundance of hot evolved stars is difficult to predict even in local star clusters and may evolve significantly with redshift (e.g., {Conroy} \& {Gunn} 2010).

In Figure \ref{fig:alternate_UVJ}, we plot 3D-HST galaxies color-coded by sSFR to contrast the performance of the UVJ diagram with FUV/MIR color-color diagrams. We optimize the dividing line in color-color space to maximize the `purity' of both quiescent and star-forming populations as sorted by their posterior median colors. Purity is defined as the fraction of galaxies in the quiescent (star-forming) box whose \texttt{Prospector}-inferred sSFRs are below (above) the target sSFR. This is done for a range of target sSFRs, and sample purity as a function of sSFR is shown below each color-color diagram. We simulate the effect of measurement uncertainty by drawing from the color posterior many times. The resulting median purities of the color-color selection are shown as dashed lines.

All of the diagrams perform fairly well at sSFR$_{\mathrm{cut}} \sim 10^{-10}$ yr$^{-1}$. However, for lower sSFRs the UVJ diagram becomes increasingly inefficient while the FUV+MIR color-color diagram remains near 100\% purity even after accounting for the effects of measurement uncertainty. This suggests that FUV/MIR fluxes are a more efficient method to select galaxies with low or very low sSFRs.

Here we report the best-fit color-color divisions for log(sSFR$_{\mathrm{cut}}$/yr$^{-1}$) = $(-9.5,-10.5,-11.5)$, respectively. Galaxies are defined as quiescent when their rest-frame colors meet the following criteria:
\begin{align}
    y &> ax+b \\
    x &> c \\
    y &> d
\end{align}
For the $V-J$, $U-V$ diagram, $a=(0.74,0.93,0.99)$, $b=(0.71,0.75,0.75)$, $c=(1.13,1.46,1.62)$, and $d=(1.6,1.46,1.31)$. 

For the $V-J$, $FUV-V$ diagram, $a=(3.24,3.84,5.03)$, $b=(0.32,0.52,0.74)$, $c=(3.45,4.98,6.14)$, and $d=(1.56,1.36,1.20)$. 

For the $V-W3$, $FUV-V$ diagram, $a=(3.12,9.17,37.73)$, $b=(-3.13,-3.62,19.40)$, $c=(2.87,4.42,5.45)$, and $d=(119.07,63.82,11.79)$.

\section{Summary and Discussion}
Here we have used Bayesian inference to show that many galaxy properties are well-correlated with their rest-frame UVJ colors. By comparing these observed correlations to fits to synthetic UVJ fluxes, we have demonstrated that correlations with M/L$_{\mathrm{g}}$, dust attenuation, and sSFR are caused by a unique mapping from colors to galaxy properties whereas correlations with stellar age and stellar metallicity are most likely driven by galaxy scaling relationships. We have used the Kullback-Leibler divergence to show that these correlations are not driven by our model priors. We have further demonstrated that the relationship between UVJ colors and sSFR begins to saturate at $\log(\mathrm{sSFR/yr}^{-1})\sim -10.5$, effectively meaning there is no sSFR-color relationship below this limit. Finally we show that the sSFR-color relationship remains robust to low levels when instead using color-color selection with FUV/MIR fluxes, and we present selection criteria in these new spaces.

First, our findings reaffirm the well-established fact that UVJ selection is largely successful in dividing the galaxy population into star-forming and quiescent systems ({Fumagalli} {et~al.} 2014). The key niche filled by the proposed new color-color diagrams is their sensitivity to sSFR below sSFR $\sim 10^{-10.5}$ yr$^{-1}$, permitting the selection of a pure sample of low-sSFR galaxies. Sample selection often involves choosing tradeoffs between purity and completeness, and optimizing for completeness produces different selection criteria which are appropriate for different science goals. A more pure quiescent sample likely will increase the efficiency of searches for high-redshift quiescent galaxies (e.g. {Schreiber} {et~al.} 2018) and may also produce cleaner distinctions between the structure of quiescent and star-forming galaxies (e.g. {Hill} {et~al.} 2019).

One challenge is that far-UV and mid-IR photometry is not always readily available. The rest-frame far-UV is easily accessible for high-redshift galaxies as it corresponds to the observed-frame UV/optical. At lower redshifts ($z\lesssim0.5$) the far-UV is only accessible through {\it GALEX}, which has lower sensitivity and angular resolution. The most robust color-color diagram requires MIR detections or upper limits. While such data are currently difficult to obtain, the upcoming launch of {\it JWST} will allow observations of the rest-frame MIR out to $z\sim1$.

This is not the first work which proposes color selection extending farther into the UV. Multiple studies find that GALEX $NUV-r$ is an excellent indicator of current versus past star formation activity {Martin} {et~al.} (2007); {Arnouts} {et~al.} (2007, 2013). {Ilbert} {et~al.} (2013) note many of the same advantages in $NUV-r$ that are found here in $FUV-V$, such as better dynamic range than $U-V$ and easier access in high-redshift galaxies. We note that tests in our framework have shown that $FUV$-based colors have a somewhat stronger correlation with sSFR than $NUV$-based colors.

These results also suggest that UVJ classification should be applied with care to spatially resolved photometry (e.g., {Liu} {et~al.} 2017). It remains to be seen whether UVJ trends which are significantly affected by galaxy scaling relationships also hold on spatially-resolved scales.

Finally we note that, while color-color diagrams are straightforward and economic choice for sample selection, more precise and accurate statements about galaxy properties can often be made by fitting models to the observed SED.

\acknowledgments
We thank Benjamin Johnson, Pieter van Dokkum, and Marijn Franx for thoughtful discussions. J.L. is supported by an NSF Astronomy and Astrophysics Postdoctoral Fellowship under award AST-1701487. S.T. is supported by the Smithsonian Astrophysical Observatory through the CfA Fellowship. The computations in this paper were run on the Odyssey cluster supported by the FAS Division of Science, Research Computing Group at Harvard University.


\end{document}